\documentclass[conference]{IEEEtran}
\IEEEoverridecommandlockouts
\usepackage{cite}
\usepackage{amsmath,amssymb,amsfonts}
\usepackage{algorithmic}
\usepackage{graphicx}
\usepackage{textcomp}
\usepackage{xcolor}
\usepackage{algorithm}
\usepackage{bm}
\usepackage[caption=false,font=footnotesize,labelfont=rm,textfont=rm]{subfig}
\def\BibTeX{{\rm B\kern-.05em{\sc i\kern-.025em b}\kern-.08em
    T\kern-.1667em\lower.7ex\hbox{E}\kern-.125emX}}
\begin{document}

\title{RIS-Aided Integrated Sensing and Communication Waveform Design With Tunable PAPR

\author{
	\IEEEauthorblockN{Jinlong Wu\IEEEauthorrefmark{1}, 
		Lixin Li\IEEEauthorrefmark{1}, 
		Wensheng Lin\IEEEauthorrefmark{1}, 
		Wei Liang\IEEEauthorrefmark{1},
		Decan Zhao\IEEEauthorrefmark{1},
		and Zhu Han\IEEEauthorrefmark{2}}
	\IEEEauthorblockA{\IEEEauthorrefmark{1}School of Electronics and Information, Northwestern Polytechnical University, Xi’an, China, 710129}
	\IEEEauthorblockA{\IEEEauthorrefmark{2}Department of Electrical and Computer Engineering, University of Houston, Houston TX, 77004}
	
\thanks{Corresponding authors: Lixin Li, Wensheng Lin.
	
	This work was supported in part by National Natural Science Foundation of China under Grant 62101450, in part by the Young Elite Scientists Sponsorship Program by the China Association for Science and Technology under Grant 2022QNRC001, in part by Aeronautical Science Foundation of China under Grants 2022Z021053001 and 2023Z071053007, in part by the Open Fund of Intelligent Control Laboratory, 
	in part by NSF CNS-2107216, CNS-2128368, CMMI-2222810, ECCS-2302469, US Department of Transportation, Toyota and Amazon.}

}

}

\maketitle

\begin{abstract}
Low peak-to-average power ratio (PAPR) transmission is an important and favorable requirement prevalent in radar and communication systems, especially in transmission links integrated with high power amplifiers. Meanwhile, motivated by the advantages of reconfigurable intelligent surface (RIS) in mitigating multi-user interference (MUI) to enhance the communication rate, this paper investigates the design problem of joint waveform and passive beamforming with PAPR constraint for integrated sensing and communication (ISAC) systems, where RIS is deployed for downlink communication. We first construct a trade-off optimization problem for the MUI and beampattern similarity under PAPR constraint. Then, in order to solve this multivariate problem, an iterative optimization algorithm based on alternating direction method of multipliers (ADMM) and manifold optimization is proposed. Finally, the simulation results show that the designed waveforms can well satisfy the PAPR requirement of the ISAC systems and achieve a trade-off between radar and communication performance. Under high signal-to-noise ratio (SNR) conditions, compared to systems without RIS, RIS-aided ISAC systems have a performance improvement of about 50\% in communication rate and at least 1 dB in beampatterning error.
\end{abstract}

\begin{IEEEkeywords}
Integrated sensing and communication, peak-to-average power ratio, reconfigurable intelligent surface, waveform design
\end{IEEEkeywords}

\section{Introduction}
Due to the potential of integrating sensing and communication functions, integrated sensing and communication (ISAC) systems have received increasing attention and are widely used in unmanned aerial vehicle (UAV) systems, intelligent transportation systems, and multi-functional radio frequency (RF) systems \cite{b1,b2}. Waveform design is a key technology for ISAC systems, and effective waveform design methods are highly required \cite{b3}. 
\par
Recently, many works are devoted to investigate ISAC waveform design to realize the trade-off between radar and communication performance. In \cite{b4} and \cite{b5}, in order to minimize the multi-user interference (MUI) while satisfying the beampattern similarity constraints and the constant-modulus constraints, the authors proposed a branch-and-bound algorithm and an algorithm based on manifold optimization to implement the ISAC waveform design, respectively. Considering the signal-dependent interference suppression problem, in \cite{b7}, the authors proposed a gradient-projection (GP) method for the joint design of transmit waveform and radar receive filters to achieve performance equalization between the signal-to-interference-plus-noise ratio (SINR) of radar signals and MUI. However, the competition between sensing performance and communication performance makes the comprehensive performance of ISAC systems limited. Recently, reconfigurable intelligent surface (RIS) have received extensive attention due to their advantages in communication systems \cite{b8,b9}. In \cite{b10}, in order to maximize the minimum signal-to-noise ratio (SNR) for users in multicast and multiuser downlink transmissions, the authors proposed a low-complexity iterative algorithm to optimize the passive beamforming vectors for RIS. In \cite{b11}, the authors considered the application of RIS in orthogonal frequency division multiple access (OFDMA) communication systems for UAVs to improve the system throughput by exploiting the beamforming gain brought by RIS and the high mobility of UAVs. In addition to the above works which concentrated on communication systems, RIS has also been used to achieve performance balance in  ISAC systems. In \cite{b12}, the authors investigated joint waveform design and passive beamforming in RIS-assisted ISAC systems and proposed an alternating algorithm based on manifold optimization to achieve the trade-off between radar and communication performance. In \cite{b13}, the authors deployed a RIS to assist downlink communications and proposed a block coordinate descent (BCD) method to maximize the SINR for radar while minimizing the MUI for communication. 
\par
The aforementioned works focus on improving the radar and communication performance of ISAC systems, while low peak-to-average power ratio (PAPR) transmission is also an important and favorable requirement for ISAC systems due to its ability to prevent nonlinear distortions caused by high power amplifiers (HPA). Therefore, low PAPR ISAC waveform design methods have also been extensively studied. In \cite{b14}, the authors proposed a semi-definite relaxation (SDR)-based method to realize ISAC waveform design under PAPR constraint. In \cite{b16} and \cite{b17}, the authors investigated the low PPAR waveform design of OFDM-based communication-sensing multiplexed integrated waveform and shared integrated waveform, respectively, and proposed an iterative optimization algorithm based on alternating direction method of multipliers (ADMM) to slove the waveform optimization problems. 
\par
In conjunction with the above works, this paper investigates the problem of designing tunable PAPR waveforms for RIS-assisted ISAC systems. We first construct a trade-off optimization problem for the communication MUI and radar beampattern similarity under PAPR constraint to achieve the joint design of ISAC waveform and RIS phase shift matrice. To solve the above optimization problem, an iterative optimization algorithm based on manifold optimization and ADMM (MO-ADMM) is proposed. The main contributions of this paper can be summarized as follows:
\begin{itemize}
	\item We propose a joint optimization framework for ISAC waveform and RIS phase shift matrix, which can achieve trade-off between radar and communication performance optimization. Meanwhile, this framework allows us to control the transmit PAPR, which is beneficial in practical physical layer design.
	\item We propose a MO-ADMM algorithm to solve the non-convex optimization problem.The ADMM is adopted for waveform optimization and the manifold optimization algorithm is adopted for RIS phase shift matrix optimization. The final optimized variables can be obtained by alternating the iterative optimization. 
	\item Simulation results show that the proposed algorithm can make the PAPR of the optimized waveform converge to the system constraint PAPR threshold. In addition, by introducing RIS, the communication performance of the ISAC system can be significantly improved, and at the same time, the designed waveform can better match the desired beampattern.
\end{itemize}
\par
$\emph{Notations:}$ $\operatorname{Re}(\cdot)$ and $\operatorname{Im}(\cdot)$ represent the real and imaginary parts of a complex variable, respectively. $(\cdot)^{T}$ and $(\cdot)^{H}$ respectively denote the transpose and conjugate transpose operations. $\left\|\cdot\right\|$ denotes the $l_2$ norm. $\operatorname{diag}(\cdot)$ is the diagonal operation. $\mathbb{E}\left( \cdot \right)$ represents the expectation of the argument.

\section{System Model}
As shown in Fig. 1, we consider an RIS-aided ISAC system consisting of one $N$-antenna base station (BS), $K$ single-antenna users, and an $L$-element RIS. The BS helps to serve $K$ users in the downlink with the help of the RIS and receives the echo signals for target detection. It is assumed that the prior location information of the detected target is konwn, which is employed to synthesize the desired beampattern.

\begin{figure}
	\centering
	\includegraphics[width=3.4in]{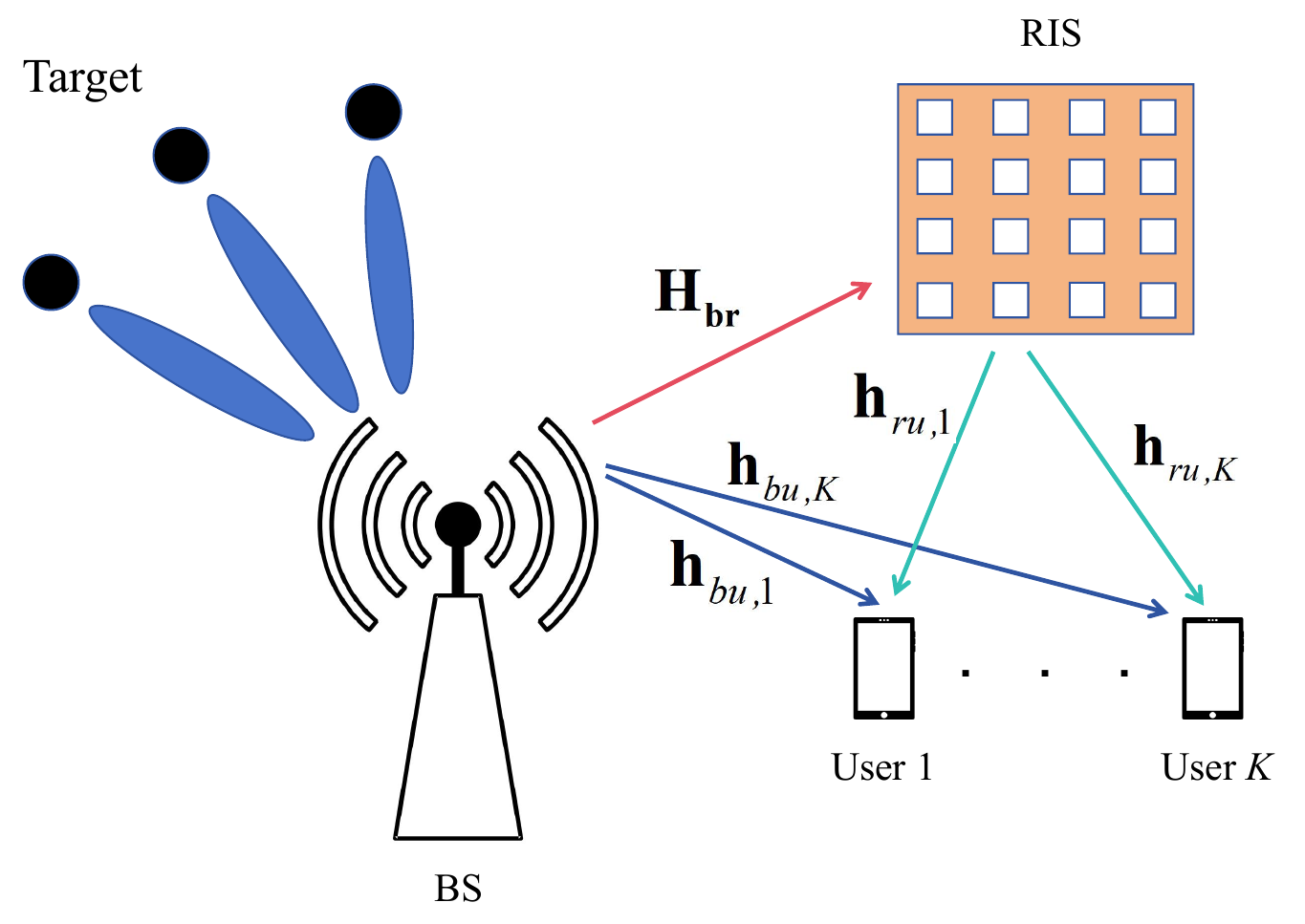}
	\caption{RIS-aided ISAC system model.}
	\label{fig1}
\end{figure}

\subsection{MIMO Communication Model}
The received signal at the users is given by
\begin{equation}\label{eq1}
	\begin{aligned}
		\mathbf{Y}=\left(\mathbf{H}_{b u}+\mathbf{H}_{r u} \mathbf{\Theta} \mathbf{H}_{b r}\right) \mathbf{X}+\mathbf{N} \triangleq \tilde{\mathbf{H}}_{b u} \mathbf{X}+\mathbf{N},
	\end{aligned}
\end{equation} 
where $\mathbf{H}_{bu}=\left[ \mathbf{h}_{bu,1},\cdots,\mathbf{h}_{bu,K} \right]^T \in \mathbb{C}^{K\times N}$, $\mathbf{H}_{ru}=\left[ \mathbf{h}_{ru,1},\cdots,\mathbf{h}_{ru,K} \right]^T \in \mathbb{C}^{K\times L}$, $\mathbf{H}_{br}\in \mathbb{C}^{L \times N} $ respectively denote the baseband channels from the BS to users, from RIS to users, and from BS to RIS. $\mathbf{\Theta} = \operatorname{diag}\left( \theta_1,\cdots,\theta_L \right)$ represent the phase shift matrix of the RIS, where $\theta_l \in \Omega$ and $\Omega=\left\{ \theta_l | |\theta_l|=1 \right\}$ is the feasible set of the reflecting coefficients of RIS. $\mathbf{X}=\left[ \mathbf{x_1},\cdots, \mathbf{x}_M \right] \in \mathbb{C}^{N \times M}$ is the transmitted waveform matrix. $\mathbf{N}=\left[ \mathbf{n}_1, \cdots, \mathbf{n}_K \right]^T \in \mathbb{C}^{K \times M} $ denotes the noise matrix at the users, where $\mathbf{n}_k \sim \mathcal{C N}\left(0, \sigma^2 \mathbf{I}\right), \forall 1 \leq k \leq K$.  $\tilde{\mathbf{H}}_{b u}=\mathbf{H}_{bu} + \mathbf{H}_{ru} \mathbf{\Theta} \mathbf{H}_{b r} $.
\par
Assuming that the desired symbol matrix at users is $\mathbf{S} \in \mathbb{C}^{K \times M}$, the received signal can be rewritten as
\begin{equation}\label{eq2}
	\begin{aligned}
		\mathbf{Y}=\mathbf{S}+\left(\tilde{\mathbf{H}}_{b u} \mathbf{X}-\mathbf{S}\right)+\mathbf{N},
	\end{aligned}
\end{equation} 
where $(\tilde{\mathbf{H}}_{b u} \mathbf{X}-\mathbf{S})$ represents the multi-user interference (MUI). The total MUI power is calculated as
\begin{equation}\label{eq3}
	\begin{aligned}
		P_{\mathrm{MUI}}=\left\|\tilde{\mathbf{H}}_{b u} \mathbf{X}-\mathbf{S}\right\|^2,
	\end{aligned}
\end{equation} 
which is closely correlated with the achievable sum-rate of the users \cite{b18}. Specifically, the signal-to-interference-plus-noise ratio (SINR) in each frame for the $k$-th user can be given as
\begin{equation}\label{eq4}
	\begin{aligned}
		\eta_k=\frac{\mathbb{E}\left(\left|s_{k, j}\right|^2\right)}{\underbrace{\mathbb{E}\left(\left|\tilde{\mathbf{h}}_{b u, k}^T \mathbf{x}_j-s_{k, j}\right|^2\right)}_{\text {MUl power }}+\sigma^2},
	\end{aligned}
\end{equation} 
where $s_{k, j}$ is the $(k,j )$-th entry of $\mathbf{S}$ and $\tilde{\mathbf{h}}_{b u, k}^T$ is the $k$-th row of $\tilde{\mathbf{H}}_{bu}$. Therefore, the achievable sum-rate of the users can be defined as
\begin{equation}\label{eq5}
	\begin{aligned}
		R=\sum_{k=1}^K \log _2\left(1+\eta_k\right)
	\end{aligned}
\end{equation} 
It can be observed that for a known constellation with fixed energy, the power of desired signal $\mathbb{E}\left(\left|s_{k, j}\right|^2\right)$ is a constant parameter. Thus, maximizing the sum-rate is almost equivalent to minimizing the MUI power. As in \cite{b4}, the MUI power is adopted as the communication performance metric.

\subsection{MIMO Radar Model}
In general, the MIMO radar beampattern design is adopted as the evaluation metric for radar waveform design, which is equivalent to the covariance matrix of the transmitted signal and can be expressed as \cite{b7}
\begin{equation}\label{eq6}
	\begin{aligned}
		P_d(\phi) = \mathbf{a}^H(\phi)\mathbf{R}_X\mathbf{a}(\phi),
	\end{aligned}
\end{equation} 
where $\mathbf{R}_X = \frac{1}{M}\mathbf{X}\mathbf{X}^H$ is the covariance matrix, $\phi$ denotes the direction angle, $\mathbf{a}(\phi)=\left[ 1,e^{j\pi sin(\phi)},\cdots,e^{j\pi(M-1)sin(\phi)} \right]$ is the steering vector. $P_d(\phi)$ reflects the spatial spectrum of the transmit waveform, and the larger the value of $P_d(\phi)$ the stronger its energy pointing to the angle in the $\phi$ direction.

\section{Waveform Design for RIS-Aided ISAC System}
In this section, we first provide the optimization problem of conventional RIS-aided ISAC waveform design under strict radar beampattern. Based on this, we propose an optimization framework with PAPR constraint for the joint design of ISAC waveform and RIS phase shift matrix. Then, the iterative algorithm MO-ADMM is proposed to  solve the joint optimization problem.
\subsection{Conventional RIS-Aided ISAC Waveform Design Under Strict Radar Beampattern}
Given a desired covariance matrix $\mathbf{R}_d$, the the RIS-aided ISAC waveform design problem can be formulated as
\begin{equation}\label{eq7}
	\begin{aligned}
		\min _{\mathbf{X}, \boldsymbol{\Theta}}  \ &\left\|\tilde{\mathbf{H}}_{b u} \mathbf{X}-\mathbf{S}\right\|_F^2 \\
		\text { s.t. }\  &\frac{1}{M} \mathbf{X} \mathbf{X}^H=\mathbf{R}_d, \\
		&\left|\theta_l\right|=1, \forall l=1, \ldots, L.
	\end{aligned}
\end{equation} 
\par
Since problem \eqref{eq7} is a multi-variate coupled problem, it needs to be decomposed into several sub-optimization problems. The above optimization problem can be solved by the method based on manifold optimization, and the specific details can be found in \cite{b12}.

\subsection{Trade-Off Joint Design With PAPR Constraint}
Since problem \eqref{eq7} needs to satisfy strict radar beampattern constraints, it leads to limited freedom in waveform design and degraded communication performance. Meanwhile, PAPR is also an important parameter for ISAC waveform design, and large PAPR value can lead to serious signal distortion in transmit waveform and affect the system performance. Therefore, we construct the following optimization problem
\begin{equation}\label{eq8}
	\begin{aligned}
		\min _{\mathbf{X}, \boldsymbol{\Theta}, \mathbf{T}} \ \ & \rho\left\|\tilde{\mathbf{H}}_{b u} \mathbf{X}-\mathbf{S}\right\|^2+(1-\rho)\|\mathbf{X}-\mathbf{T}\|^2, \\
		\text { s.t. }\ \  & \frac{1}{M} \mathbf{T T}^H=\mathbf{R}_d, \\
		& \|\mathbf{X}\|^2=M P_t, \\
		& \left|\theta_l\right|=1, \forall l=1, \cdots, L, \\
		& \operatorname{PAPR}(\mathbf{X}) \leq \eta,
	\end{aligned}
\end{equation} 
where $\rho$ is the weighting factor that balances radar and communication performances. $\mathbf{T}$ is the waveform template used to match the desired radar beampattern and its initial value can be obtained from problem \eqref{eq7}. $P_t$ is the total transmitted power. $\eta$ denotes the PAPR threshold of the waveform. 

\subsection{The Proposed MO-ADMM Algorithm}
To solve the non-convex problem \eqref{eq8}, we propose an iterative algorithm MO-ADMM to iteratively optimize each variable while fixing the other variables.
\par
\emph{1) Optimization $\mathbf{X}$ for Given $\mathbf{\Theta}$ and $\mathbf{T}$}:
Assuming $\mathbf{A}=\left[\sqrt{\rho} \tilde{\mathbf{H}}_{b u}^T, \sqrt{1-\rho} \mathbf{I}_N\right]^T, \mathbf{B}=\left[\sqrt{\rho} \mathbf{S}^T, \sqrt{1-\rho} \mathbf{T}^T\right]^T$, the objective function of equation \eqref{eq8} can be expressed as $\|\mathbf{A X}-\mathbf{B}\|^2$. The PAPR constraint of equation \eqref{eq8} can be rewritten as
\begin{equation}\label{eq9}
	\begin{aligned}
		\operatorname{PAPR}(\mathbf{x}) = \frac{\mathop{\max}\limits_{m}  |x(m)|^2}{\frac{\|\mathbf{x}\|^2}{NM} } \leq \eta,
	\end{aligned}
\end{equation} 
where $\mathbf{x}=\operatorname{vec}(\mathbf{X}) \in \mathbb{C}^{NM\times 1}$. It can be observed that the total transmit power and PAPR constraints can be converted to quadratic equality and inequality constraints, respectively, which can be expressed as
 \begin{equation}\label{eq10}
 	\begin{aligned}
 		\mathbf{x}^H \mathbf{x}=M P_t,
 	\end{aligned}
 \end{equation} 
 \begin{equation}\label{eq11}
 	\begin{aligned}
 		\mathbf{x}^H \mathbf{E}_m \mathbf{x} \leq \frac{P_t \eta}{N},\ m = 1,\cdots,NM
 	\end{aligned}
 \end{equation} 
where
\begin{equation}\label{eq12}
	\begin{aligned}
		\mathbf{E}_m(i, j)= \begin{cases}1, & i=m ,\  j=m \\ 0, & \text { otherwise. }\end{cases}
	\end{aligned}
\end{equation} 
\par
To combine the vectorization constraints with the objective function, a diagonal matrix $\tilde{\mathbf{A}}$ can be defined as
\begin{equation}\label{eq13}
	\begin{aligned}
		\tilde{\mathbf{A}}=\left[\begin{array}{lllll}
			\mathbf{A} & & & & \\
			& \mathbf{A} & & 0 & \\
			& & \ddots & & \\
			& 0 & & \mathbf{A} & \\
			& & & & \mathbf{A}
		\end{array}\right] \in \mathbb{C}^{(K+N) M \times N M}.
	\end{aligned}
\end{equation} 
\par
Then, ignoring the terms not related to $\mathbf{x}$, equation \eqref{eq8} can be rewritten as
\begin{equation}\label{eq14}
	\begin{aligned}
		\min _{\mathbf{x}} \  &  \|\tilde{\mathbf{A}} \mathbf{x}-\mathbf{b}\|^2 ,\\
		\text { s.t. } \  &  \mathbf{x}^H \mathbf{x}=M P_{\mathrm{t}}, \\
		& \mathbf{x}^H \mathbf{E}_m \mathbf{x} \leq \frac{P_{\mathrm{t}} \eta}{N}, \ m=1,\cdots,NM
	\end{aligned}
\end{equation} 
where $\mathbf{b}=\operatorname{vec}(\mathbf{B}) \in \mathbb{C}^{(K+N) M \times 1}$. Problem \eqref{eq14} is a non-convex quadratically constrained quadratic programs (QCQP) problem and semi-definite relaxation (SDR)  techniques can be used to solve such problems, but its computational complexity is relatively high. Here, we adopt the ADMM algorithm \cite{b19} to solve problem \eqref{eq14} by decomposing it into multiple sub problems and then iteratively optimizing them alternately.
\par
Firstly, convert the complex variables in equation \eqref{eq14} into real variables, we can obtain
\begin{equation}\label{eq15}
	\begin{aligned}
		\overline{\mathbf{A}} =\left[\begin{array}{cc}
			\operatorname{Re}(\tilde{\mathbf{A}}) & -\operatorname{Im}(\tilde{\mathbf{A}}) \\
			\operatorname{Im}(\tilde{\mathbf{A}}) & \operatorname{Re}(\tilde{\mathbf{A}})
		\end{array}\right],
	\end{aligned}
\end{equation} 
\begin{equation}\label{eq16}
	\begin{aligned}
		\overline{\mathbf{x}}=\left[\begin{array}{c}
			\operatorname{Re}(\mathbf{x}) \\
			\operatorname{Im}(\mathbf{x})
		\end{array}\right],
	\end{aligned}
\end{equation} 
\begin{equation}\label{eq17}
	\begin{aligned}
		\overline{\mathbf{b}}=\left[\begin{array}{c}
			\operatorname{Re}(\mathbf{b}) \\
			\operatorname{Im}(\mathbf{b})
		\end{array}\right].
	\end{aligned}
\end{equation} 
\par
Then, introducing auxiliary variables $\boldsymbol{\alpha}$ and $\boldsymbol{\gamma}_n,n=1,\cdots,NM$, such that $\boldsymbol{\alpha}=\overline{\mathbf{x}}$, $\boldsymbol{\gamma}_n=\overline{\mathbf{E}}_n\overline{\mathbf{x}}$, where
\begin{equation}\label{eq18}
	\begin{aligned}
		\overline{\mathbf{E}}_n(i, j)=\left\{\begin{array}{lc}
			1, & i=n, j=n \\
			1, & i=n+N M, j=n+N M . \\
			0, & \text { otherwise }
		\end{array}\right.
	\end{aligned}
\end{equation} 
\par
Therefore, \eqref{eq14} can be rewritten as
\begin{equation}\label{eq19}
	\begin{aligned}
		\min _{\mathbf{x}} \  &  \|\tilde{\mathbf{A}} \mathbf{x}-\mathbf{b}\|^2 ,\\
		\text { s.t. } \  & \overline{\mathbf{x}}=\boldsymbol{\alpha}, \\
		&  \boldsymbol{\alpha}^T \boldsymbol{\alpha}=M P_{\mathrm{t}}, \\
		&  \overline{\mathbf{E}}_n\overline{\mathbf{x}} = \boldsymbol{\gamma}_n, \ n=1,\cdots,NM, \\
		& \boldsymbol{\gamma}_n^T \boldsymbol{\gamma}_n \leq \frac{P_{\mathrm{t}} \eta}{N}, \ n=1,\cdots,NM.
	\end{aligned}
\end{equation} 
\par
According to \eqref{eq19}, the augmented Lagrangian function can be described as
\begin{equation}\label{eq20}
	\begin{aligned}
		L= & \|\overline{\mathbf{A}} \overline{\mathbf{x}}-\overline{\mathbf{b}}\|^2+\mathbf{u}^T(\overline{\mathbf{x}}-\boldsymbol{\alpha})+\sum_{n=1}^{N M} \mathbf{w}_n^T\left(\overline{\mathbf{E}}_n \overline{\mathbf{x}}-\boldsymbol{\gamma}_n\right) \\
		& +\frac{\mu}{2}\|\overline{\mathbf{x}}-\boldsymbol{\alpha}\|^2+\frac{\mu}{2} \sum_{n=1}^{N M}\left\|\overline{\mathbf{E}}_n \overline{\mathbf{x}}-\boldsymbol{\gamma}_n\right\|^2
	\end{aligned}
\end{equation} 
where $\mathbf{u}$ and $\mathbf{w}_n$ are the Lagrange multiplier vector, and $\mu > 0$ denotes the penalization factor. The parameters will be updated by alternating iteration. In the $(m+1)$-th iteration, the parameters will be updated as follows
\begin{equation}\label{eq21}
	\begin{aligned}
		\overline{\mathbf{x}}^{(m+1)}=\arg \min _{\overline{\mathbf{x}}} \ L\left(\overline{\mathbf{x}},[\boldsymbol{\alpha}, \boldsymbol{\gamma}, \mathbf{u}, \mathbf{w}]^{(m)}\right)
	\end{aligned}
\end{equation} 

\begin{equation}\label{eq22}
	\begin{aligned}
		\boldsymbol{\alpha}^{(m+1)}=\arg \min _{\boldsymbol{\alpha}} \  L\left(\overline{\mathbf{x}}^{(m+1)}, \boldsymbol{\alpha},[\boldsymbol{\gamma}, \mathbf{u}, \mathbf{w}]^{(m)}\right)
	\end{aligned}
\end{equation} 

\begin{equation}\label{eq23}
	\begin{aligned}
		\boldsymbol{\gamma}^{(m+1)}=\arg \min _{\boldsymbol{\gamma}} \ L\left([\overline{\mathbf{x}}, \boldsymbol{\alpha}]^{(m+1)}, \boldsymbol{\gamma},[\mathbf{u}, \mathbf{w}]^{(m)}\right)
	\end{aligned}
\end{equation} 

\begin{equation}\label{eq35}
	\begin{aligned}
		\mathbf{u}^{(m+1)} = \mathbf{u}^{(m)} + \mu\left( \overline{\mathbf{x}}-\boldsymbol{\alpha} \right)
	\end{aligned}
\end{equation} 

\begin{equation}\label{eq36}
	\begin{aligned}
		\mathbf{w}_n^{(m+1)} = \mathbf{w}_n^{(m)} + \mu\left( \overline{\mathbf{E}}_n\overline{\mathbf{x}}-\boldsymbol{\gamma}_n \right)
	\end{aligned}
\end{equation} 

\par
The specific solution procedures for problems \eqref{eq21}, \eqref{eq22} and \eqref{eq23} will be given below,  and the superscripts are omitted for brevity.
\par
$\emph{Step 1:}$ Updating $\overline{\mathbf{x}}$ for given $\left\{ {\boldsymbol{\alpha}}, \boldsymbol{\gamma}, \mathbf{u}, \mathbf{w} \right\}$
\par
According to \eqref{eq21}, by solving the gradient of the augmented Lagrangian function with respect to $\overline{\mathbf{x}}$ and setting it to zero, i.e.
\begin{equation}\label{eq24}
	\begin{aligned}
		\nabla_{\overline{\mathbf{x}}}\left[L\left(\overline{\mathbf{x}}, \boldsymbol{\alpha}, \boldsymbol{\gamma}, \mathbf{u}, \mathbf{w}\right)\right] = \mathbf{0}, 
	\end{aligned}
\end{equation} 
we can get
\begin{equation}\label{eq25}
	\begin{aligned}
		2 \overline{\mathbf{A}}^T \overline{\mathbf{A}} \overline{\mathbf{x}}-2 \overline{\mathbf{A}} \overline{\mathbf{b}}+\mathbf{u}+\sum_{n=1}^{N M} \overline{\mathbf{E}}_n^T \mathbf{w}_n
		+\mu(\overline{\mathbf{x}}-\boldsymbol{\alpha}) \\
		+\mu \sum_{n=1}^{N M}\left(\overline{\mathbf{E}}_n^T \overline{\mathbf{E}} \overline{\mathbf{x}}_n \overline{\mathbf{E}}_n^T \boldsymbol{\gamma}_n\right)=\mathbf{0}.
	\end{aligned}
\end{equation} 
\par
Utilizing the symmetry and idempotency properties of $\overline{\mathbf{E}}_n$, i.e. $\overline{\mathbf{E}}_n^T = \overline{\mathbf{E}}_n$ and $\overline{\mathbf{E}}_n \overline{\mathbf{E}}_n = \overline{\mathbf{E}}_n$, and rearranging \eqref{eq25}, the update equation for $\overline{\mathbf{x}}$ can be written as
\begin{equation}\label{eq26}
	\begin{aligned}
		\overline{\mathbf{x}}=\left(2 \overline{\mathbf{A}}^T \overline{\mathbf{A}}+2 \mu \mathbf{I}_{2 N M}\right)^{+}\left(2 \overline{\mathbf{A}} \overline{\mathbf{b}}-\mathbf{u}-\sum_{n=1}^{N M} \overline{\mathbf{E}}_n^T \mathbf{w}_n \right. \\
 		\left. +\mu \boldsymbol{\alpha}+\mu \sum_{n=1}^{N M} \overline{\mathbf{E}}_n^T \boldsymbol{\gamma}_n\right)
	\end{aligned}
\end{equation} 
\par
$\emph{Step 2:}$ Updating $\boldsymbol{\alpha}$ for given $\left\{ \overline{\mathbf{x}}, \boldsymbol{\gamma}, \mathbf{u}, \mathbf{w} \right\}$
\par
Ignoring the irrelevant terms to $\boldsymbol{\alpha}$, \eqref{eq22} can be simplified as
\begin{equation}\label{eq27}
	\begin{aligned}
		\min _{\boldsymbol{\alpha}} & \ \mathbf{u}^T(\overline{\mathbf{x}}-\boldsymbol{\alpha})+\frac{\mu}{2}\|\overline{\mathbf{x}}-\boldsymbol{\alpha}\|^2 \\
		\text { s.t. } & \ \boldsymbol{\alpha}^T \boldsymbol{\alpha}=M P_t,
	\end{aligned}
\end{equation}
whose Lagrangian function can be written as
\begin{equation}\label{eq28}
	\begin{aligned}
		L_\alpha(\boldsymbol{\alpha}, \lambda)=-\mathbf{u}^T \boldsymbol{\alpha}+\frac{\mu}{2}\left(\boldsymbol{\alpha}^T \boldsymbol{\alpha}-2 \boldsymbol{a}^T \bar{x}\right)+\lambda\left(\boldsymbol{\alpha}^T \boldsymbol{\alpha}-M P_t\right) .
	\end{aligned}
\end{equation}
\par
Letting the gradient of $L_\alpha$ with respect to $\boldsymbol{\alpha}$ be zero, i.e.
\begin{equation}\label{eq29}
	\begin{aligned}
		\nabla_{\boldsymbol{\alpha}} \left[L_\alpha(\boldsymbol{\alpha}, \lambda)\right]=\mathbf{0},
	\end{aligned}
\end{equation}
we have
\begin{equation}\label{eq30}
	\begin{aligned}
		-\mathbf{u}+\mu(\boldsymbol{\alpha}-\overline{\mathbf{x}})+2 \lambda \boldsymbol{\alpha}=\mathbf{0},
	\end{aligned}
\end{equation}
which can be further expressed as
\begin{equation}\label{eq31}
	\begin{aligned}
		\boldsymbol{\alpha}=\frac{1}{\mu+2 \lambda}(\mathbf{u}+\mu \overline{\mathbf{x}}).
	\end{aligned}
\end{equation}
\par
In general, we can set the Lagrange multiplier $\lambda$ to zero and normalize the vector $\boldsymbol{\alpha}$ to satisfy the constraint $\boldsymbol{\alpha}^T\boldsymbol{\alpha}=MP_t$. Thus, \eqref{eq31} can be rewritten as
\begin{equation}\label{eq32}
	\begin{aligned}
		\boldsymbol{a}=\sqrt{M P_t} \frac{\overline{\mathbf{x}}+\frac{\mathbf{u}}{\mu}}{\left\|\overline{\mathbf{x}}+\frac{\mathbf{u}}{\mu}\right\|} .
	\end{aligned}
\end{equation}
\par
$\emph{Step 3:}$ Updating $\boldsymbol{\gamma}_n$ for given $\left\{ \overline{\mathbf{x}}, \boldsymbol{\alpha}, \mathbf{u}, \mathbf{w} \right\}$
\par
Ignoring the irrelevant terms to $\boldsymbol{\gamma}_n$, \eqref{eq23} can be simplified as
\begin{equation}\label{eq33}
	\begin{aligned}
		\min _{\gamma_n} &\ \mathbf{w}_n^T\left(\overline{\mathbf{E}}_n \overline{\mathbf{x}}-\boldsymbol{\gamma}_n\right)+\frac{\mu}{2}\left\|\overline{\mathbf{E}}_n \overline{\mathbf{x}}-\boldsymbol{\gamma}_n\right\|^2 \\
		\text { s.t. } & \ \boldsymbol{\gamma}_n^T\boldsymbol{\gamma}_n \leq \frac{P_t \eta}{N}, n = 1,\cdots, NM.
	\end{aligned}
\end{equation}
\par
The solution of \eqref{eq33} can be given as
\begin{equation}\label{eq34}
	\begin{aligned}
		\boldsymbol{\gamma}_n=\left\{\begin{array}{ll}
			\overline{\mathbf{E}}_n \overline{\mathbf{x}}+\frac{1}{\mu} \mathbf{w}_n, & \boldsymbol{\gamma}_n^T \boldsymbol{\gamma}_n \leq \frac{P_t \eta}{N} \\
			\sqrt{\frac{P_t \eta}{N}} \frac{\overline{\mathbf{E}}_n \overline{\mathbf{x}}+\frac{1}{\mu} \mathbf{w}_n}{\left\|\overline{\mathbf{E}}_n \overline{\mathbf{x}}+\frac{1}{\mu} \mathbf{w}_n\right\|}, & \text {otherwise}
		\end{array} .\right.
	\end{aligned}
\end{equation}
\par
By alternately iterating and optimizing the above steps, the optimized waveform $\mathbf{X}$ can be obtained according to \eqref{eq26}.

\emph{2) Optimization $\mathbf{T}$ for given $\mathbf{X}$ and $\mathbf{\Theta}$}:
\par
Ignoring the irrelevant terms to $\mathbf{T}$, \eqref{eq8} can be simplified as
\begin{equation}\label{eq37}
	\begin{aligned}
		\min _{\mathbf{T}} & \  \|\mathbf{T}-\mathbf{X}\|^2 \\
		\text{s.t.} & \ \frac{1}{M}\mathbf{T}\mathbf{T}^T = \mathbf{R}_d,
	\end{aligned}
\end{equation}
\par
Defining the Cholesky decomposition of $\mathbf{R}_d$ as $\mathbf{R}_d = \mathbf{F}\mathbf{F}^H$, where $\mathbf{F} \in \mathbb{C}^{N \times N}$ is a lower triangular matrix. According to \cite{b4}, the optimal solution of problem \eqref{eq37} is described as
\begin{equation}\label{eq38}
	\begin{aligned}
		\mathbf{T} = \sqrt{M}\mathbf{F}\overline{\mathbf{U}}\mathbf{I}_{N \times M}\overline{\mathbf{V}}^H,
	\end{aligned}
\end{equation}
where $\overline{\mathbf{U}} \bar{\Sigma} \overline{\mathbf{V}}^H=\mathbf{F}^H \mathbf{X}$ is the singular value decomposition (SVD) of $\mathbf{F}^H\mathbf{X}$.

\emph{3) Optimization $\mathbf{\Theta}$ for given $\mathbf{X}$ and $\mathbf{T}$}:
\par
Ignoring the irrelevant terms to $\mathbf{\Theta}$, and bringing $\tilde{\mathbf{H}}_{b u}=\mathbf{H}_{b u}+\mathbf{H}_{r u} \boldsymbol{\Theta} \mathbf{H}_{b r}$ into \eqref{eq8}, we can get
\begin{equation}\label{eq39}
	\begin{aligned}
		\min _{\boldsymbol{\theta}} & \  \operatorname{Tr}\left(\mathbf{\Theta}^H \mathbf{B \Theta C}\right)+\operatorname{Tr}\left(\mathbf{\Theta}^H \mathbf{D}^H\right)+\operatorname{Tr}(\mathbf{\Theta D}) \\
		\text {s.t. } & \  \left|\theta_l\right|=1, \forall l=1, \cdots, L
	\end{aligned}
\end{equation}
where $\mathbf{B} = \mathbf{H}_{ru}^H\mathbf{H}_{ru}$, $\mathbf{C} = \mathbf{H}_{br}\mathbf{X}\mathbf{X}^H\mathbf{H}_{br}^H$, $\mathbf{D} = \mathbf{H}_{br}\mathbf{X}\left(\mathbf{H}_{bu}\mathbf{X}-\mathbf{S}\right)^H\mathbf{H}_{ru}$. Further, defining $\boldsymbol{\theta} = \mathbf{\Theta}\mathbf{1}_L$ and $\mathbf{d} = \left[ D_{1,1},\cdots,D_{L,L} \right]^T$, where $D_{l,l}$ denotes the $l$-th diagonal element of the matrix $\mathbf{D}$, yields
\begin{equation}\label{eq40}
	\begin{aligned}
		\min _\theta & \ \boldsymbol{\theta}^H\left(\mathbf{B} \odot \mathbf{C}^T\right) \boldsymbol{\theta}+\mathbf{d}^T \boldsymbol{\theta}+\boldsymbol{\theta}^H \mathbf{d}^* \\
		\text {s.t. } & \  \left|\theta_l\right|=1, \forall l=1, \cdots, L,
	\end{aligned}
\end{equation}
which is a non-convex modulus constrained problem. According to \cite{b12}, the manifold optimization algorithm can be used to solve it, which can be obtained as
\begin{equation}\label{eq41}
	\begin{aligned}
		\boldsymbol{\theta}=R_{\theta}\left(\alpha \xi\right),
	\end{aligned}
\end{equation}
where $R_{\theta}$ denotes a retraction of $\boldsymbol{\theta}$ on the manifold, $\alpha$ is the step size, and $\xi$ is the negative counterpart of the Riemannian gradient. 
\par
Based on the above analysis, the proposed MO-ADMM algorithm for problem \eqref{eq8} is summarized in Algorithm 1.
\begin{algorithm}
	\caption{The proposed MO-ADMM algorithm for solving problem \eqref{eq8}.}\label{alg1}
	\begin{algorithmic}
		\STATE
		\STATE $\textbf{Initialization:}$ $\mathbf{X}^{(0)}, \boldsymbol{\alpha}^{(0)}, \boldsymbol{\gamma}_n^{(0)}, \mathbf{u}^{(0)}, \mathbf{w}_n^{(0)}, \mathbf{\Theta}^{(0)}$, 
		 maximum iteration number $M_{iter}$ and set $m=0$.
		\STATE $ \textbf{Repeat} $
		\STATE \hspace{0.5cm} Obtain $ \mathbf{X}^{(m+1)} $ using \eqref{eq26}.
		\STATE \hspace{0.5cm} Obtain $ \mathbf{T}^{(m+1)} $ using \eqref{eq38}.
		\STATE \hspace{0.5cm} Obtain $ \mathbf{\Theta}^{(m+1)} $ using \eqref{eq41}.
		\STATE \hspace{0.5cm} Update $ m = m+1 $.
		\STATE $\textbf{Until}$ $m=M_{iter}$ or the value of objective function converged. 
		\STATE $\textbf{Output:}$ the optimized waveform $\mathbf{X}$.
	\end{algorithmic}
\end{algorithm}
\par
Meanwhile, we also analyze the computational complexity of Algorithm 1. The complexity of optimization $\mathbf{X}$ for given $\mathbf{\Theta}$ and $\mathbf{T}$ is $O(N^3M^3+KN^2M^3)$. The complexity of optimization $\mathbf{T}$ for given $\mathbf{\Theta}$ and $\mathbf{X}$ is $O(NM^2+N^2M+NKM+N^3+N^2K)$. The complexity of optimization $\mathbf{\Theta}$ for given $\mathbf{T}$ and $\mathbf{X}$ is $O(N^2L)$. Hence, the total complexity of the proposed method is $O(N^3M^3+KN^2M^3+N^2L)$ per iteration.

\begin{figure}
	\centering
	\includegraphics[width=3.4in]{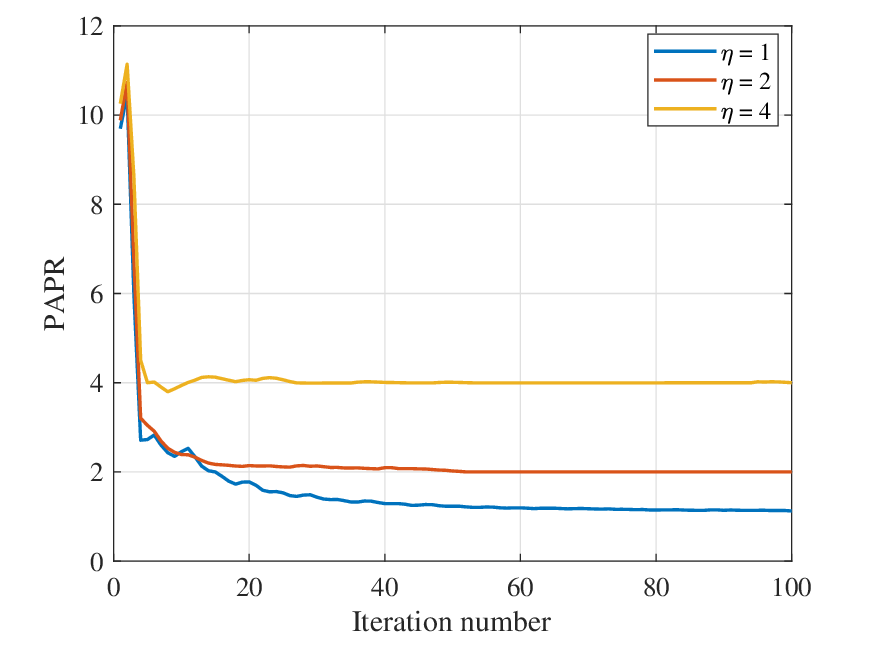}
	\caption{The PAPR convergence behavior of the proposed MO-ADMM algorithm per iteration for different values of $\eta$.}
	\label{fig2}
\end{figure}

\section{Simulation Results}
In this section, numerical simulations are implemented to demonstrate the performance of the proposed MO-ADMM algorithm. According to the assumption in \cite{b4}, each entry of the channel matrix subject to standard Complex Gaussian distribution. The simulation experiments are performed with the number of antennas $N=16$ at the BS, the total signal transmission power $P_t = 20$ dBm, the number of users $K=4$, the number of RIS elements $L=20$, and the communication frame length $M=20$. And, it is assumed that each of the three targets are located at the angles of $\left[ -45^{\circ} , 0^{\circ}, 45^{\circ} \right]$ from the BS and the angle information is known a priori. We also compared the results with no aid of RIS, which is denoted as `w/o RIS', and `w RIS' indicates with the aid of RIS.
\par
Fig. 2 gives the PAPR convergence behavior of the proposed MO-ADMM algorithm per iteration for different values of $\eta$. From Fig. 2, we can see that the proposed algorithm is enabled to make the waveform PAPR converge to the specified $\eta$ value. Moreover, it can be observed that the number of iterations required for PAPR convergence depends on the value of $\eta$. The smaller the value of $\eta$, the larger the number of iterations.
\par
Fig. 3 shows the sum rates achievable by the proposed MO-ADMM algorithm for different transmit SNRs, where $\rho=0.1, \eta=2$, and the ADMM and SDR algorithm are considered as the benchmark algorithms. From Fig. 3, we can see that the sum rate achievable rises with the increase of SNR. Moreover, compared to the `w/o RIS' systems, the achievable sum rate of the `w RIS' system can be remarkably improved.
\par
Fig. 4 demonstrates the resultant radar beampatterns for different algorithms. From Fig. 4(a), it can be observed that the radar beampattern achieved by the proposed MO-ADMM algorithm can be better matched with the desired radar beampattern compared to the ADMM and SDR algorithms. This is due to the fact that RIS can mitigate the MUI making the optimization problem approximate to minimize the beampattern matching error term. In order to display the difference between the designed waveform beampattern and the desired beampattern moe clearly, Fig. 4(b) shows the mean square error (MSE) of the designed waveform beampatterns by different algorithms versus the weighting factor $\rho$. As can be seen, compared with the `w/o RIS' systems, the MSE of `w RIS' system can be significantly reduced. This results indicate that RIS can better balance the radar and communication performance of the ISAC system.



\begin{figure}
	\centering
	\includegraphics[width=3.4in]{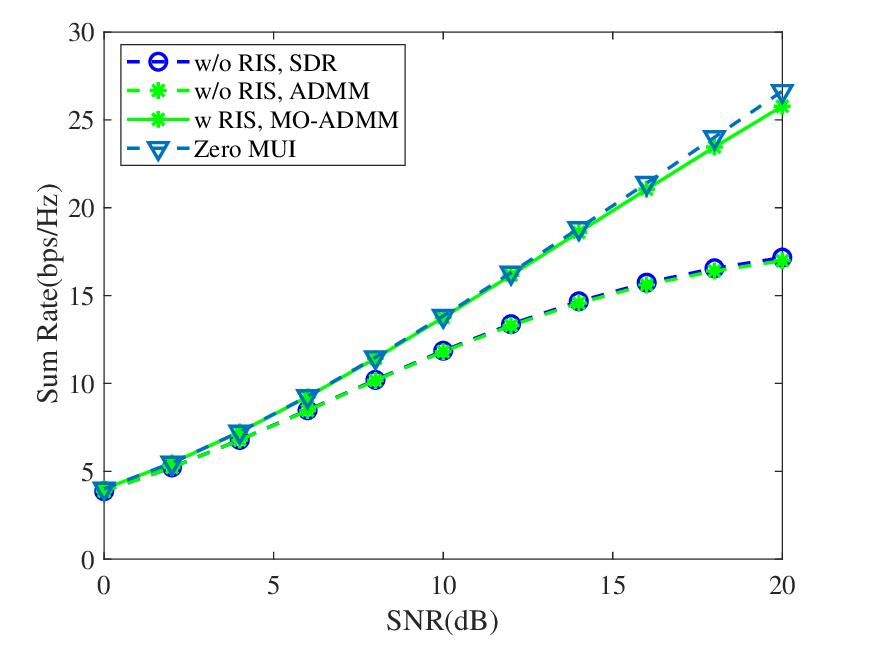}
	\caption{The sum rate versus the SNR.}
	\label{fig3}
\end{figure}

\begin{figure}
	\centering
	\vspace{-0.4cm}
	\subfloat[\centering{}]{\includegraphics[width = 0.5\linewidth]{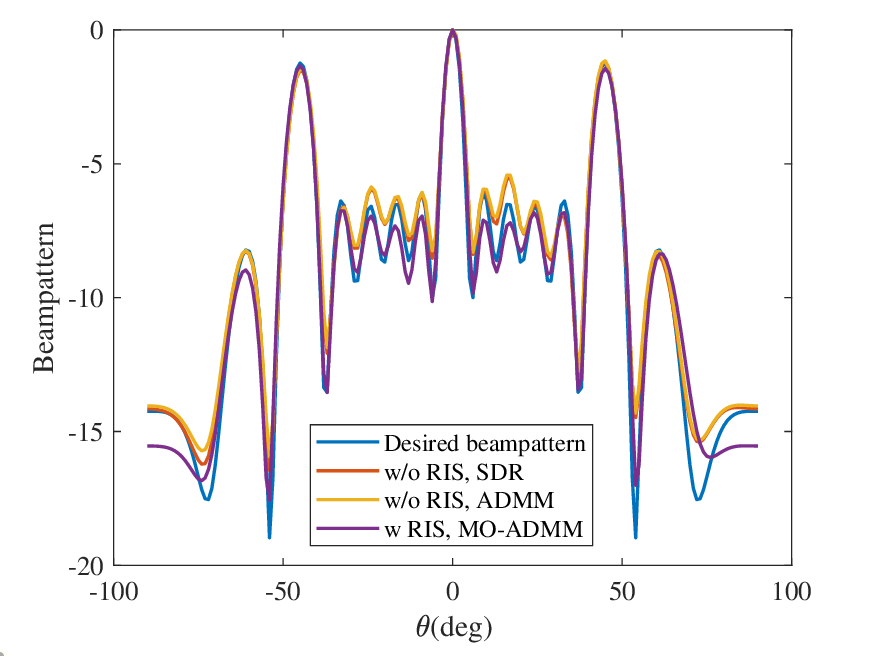}}
	\hfill
	\subfloat[\centering{}]{\includegraphics[width = 0.5\linewidth]{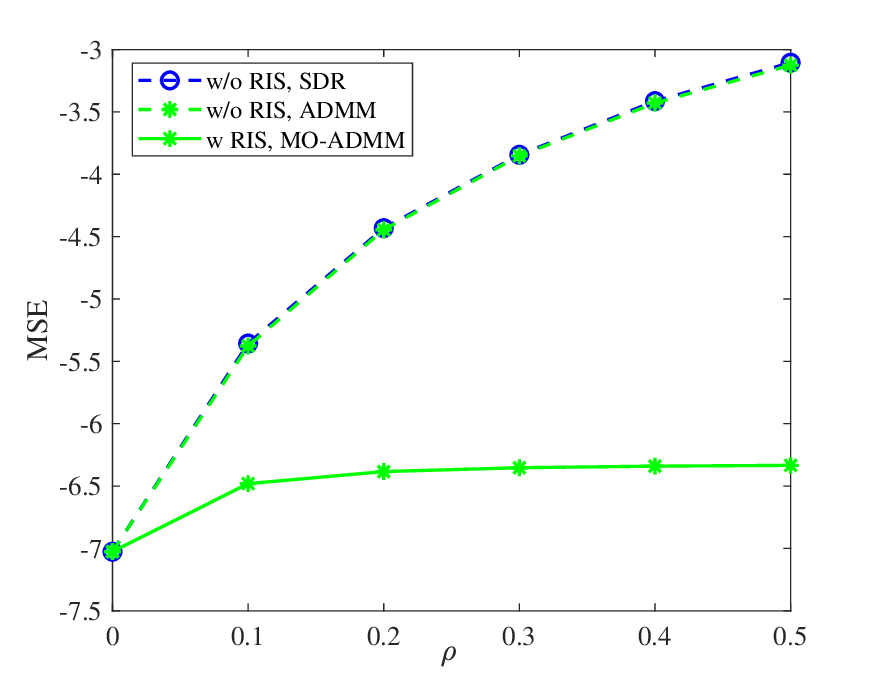}}
	\caption{(a) Radar beampattern of different methods. (b) MSE of the obtained beampatterns by different methods versus weighting factor $\rho$.}
	\label{fig6}
\end{figure}

\section{Conclusion}
In this paper, we have investigated joint waveform design and passive beamforming for RIS-aided ISAC system. In order to minimize the trade-off for MUI and beampattern similarity with the total transmit power and PAPR constraints, the MO-ADMM algorithm is proposed, which decomposes the multivariate coupled optimization problem and solves it through alternating iterative optimization. The simulation results show that the PAPR value of the designed waveforms by proposed MO-ADMM algorithm can converge to the required PAPR value of the system. Moreover, simulation results also verify the advantages of RIS in mitigating MUI and balancing communication and radar performances.

\end{document}